\documentclass[letterpaper,english,aps,prb,floatfix,showpacs,amsmath,amssymb,eqsecnum,superscriptaddress,twocolumn]{revtex4-1}
\usepackage[T1]{fontenc}
\usepackage[latin9]{inputenc}
\usepackage{color}
\usepackage{babel}
\usepackage{amsmath}
\usepackage{amssymb}
\usepackage{graphicx}
\usepackage[unicode=true,pdfusetitle,
 bookmarks=true,bookmarksnumbered=false,bookmarksopen=false,
 breaklinks=false,pdfborder={0 0 1},backref=false,colorlinks=false]
 {hyperref}

\makeatletter

\pdfpageheight\paperheight
\pdfpagewidth\paperwidth

\makeatother

\begin{document}
\title{Calculating DMFT forces in \foreignlanguage{british}{\textit{ab-initio}}
ultrasoft pseudopotential formalism}
\author{Evgeny Plekhanov}
\email{evgeny.plekhanov@kcl.ac.uk}

\affiliation{King's College London, Theory and Simulation of Condensed Matter (TSCM),
The Strand, London WC2R 2LS, United Kingdom}

\author{Nicola Bonini}
\affiliation{King's College London, Theory and Simulation of Condensed Matter (TSCM),
The Strand, London WC2R 2LS, United Kingdom}

\author{Cedric Weber}
\email{cedric.weber@kcl.ac.uk}
\affiliation{King's College London, Theory and Simulation of Condensed Matter (TSCM),
The Strand, London WC2R 2LS, United Kingdom}
\thanks{A footnote to the article title}
\date{\today}
\begin{abstract}
In this paper, we show how to calculate analytical atomic forces within
self-consistent density functional theory + dynamical mean-field theory (DFT+DMFT) approach in the case when ultrasoft or norm-conserving pseudopotentials
are used. We show how to treat the non-local projection terms arising
within the pseudopotential formalism and circumvent the problem of non-orthogonality of the Kohn-Sham eigenvectors.
Our approach 
is, in principle, independent of the DMFT solver employed, and here was tested with the Hubbard I solver. 
We benchmark our formalism by
comparing against the forces calculated in Ce$_{2}$O$_{3}$ and PrO$_2$ by numerical
differentiation of the total free energy as well as by comparing the energy
profiles against the numerically integrated analytical forces.
\end{abstract}
\maketitle

\section{Introduction}

The ability to calculate atomic forces in quantum systems allows
for efficient exploration of the energy landscape. This, in turns,
stays at the origin of several crucial approaches in condensed matter
physics: structural optimization, new material design, molecular dynamics
and so on. Within density functional theory (DFT), the calculation
of forces is based on the variational properties of the DFT total
energy functional on one hand and on the Hellmann-Feynman theorem
on the other. As a result, the forces within the all-electron DFT
can be calculated based on the explicit dependence of the ion-ion
and the ion-electron interaction terms on the atomic positions.

On the other hand, practical DFT calculations rely on approximate
exchange-correlation functionals, which handicaps the ability of DFT
to reproduce strongly correlated physics in many materials, notably
those containing open $d$ or $f$-shell elements. Many strongly-correlated
materials exhibit properties useful for technological applications\citep{Kotliar_Vollhardt_2004,Weber_2012,Weber_2013}.
For example, the copper oxides and iron pnictides are high temperature
superconductors\citep{Plekhanov_2003,Plekhanov_2005,Dai_2015}, and
the cobaltates exhibit colossal thermoelectric power\citep{Lija_2014}
which is useful for energy conversion. Several vanadates have peculiar
room-temperature metal-insulator transitions, allowing realization
of a so-called \textquotedblleft intelligent window\textquotedblright ,
which becomes insulating as the external temperature drops\citep{Babulanam_1987,Granqvist_1990,Granqvist_2007,Tomczak_2009}. 

The failure of DFT's exchange-correlation functionals to capture strong
correlation physics severely limits its use for nano-scale design
of such important functional materials. In contrast
to DFT, huge progress has been made in describing strongly-correlated
materials with Dynamical Mean-Field Theory (DMFT)\citep{Georges_1996,Vollhardt_2010,Savrasov_2004,Minar_2005,Kotliar_2006,Pourovskii2007,Amadon_2008,Amadon2012,Monserrat_2020}.
DMFT is a sophisticated method which offers a higher level of theoretical
description than DFT, and bridges the gap between DFT and Green function
approaches. Within DMFT, the treatment of local electronic correlation
effects is formally exact, although the non-local electronic correlation
effects are neglected. DMFT can be combined with DFT giving rise
to the DFT+DMFT method\citep{Savrasov_2004,Kotliar_2006,Pourovskii2007,CASTEPDMFT,CTQMCCASTEP,PaperLantanides,PaperY2020},
where the DMFT is applied to selected ``correlated'' $d$ and/or
$f$ orbitals, while the rest of the system is treated at the DFT
level. Moreover, within DFT+DMFT, a variational principle for the
total free energy can be derived\citep{Kotliar_2006,Haule_2015_Free_Energy}
and it can be shown that, at self-consistent DFT+DMFT solution corresponds
to a stationary point.

There has been several approaches to the calculation of forces within
DFT+DMFT. In the work of Savrasov \emph{et al.}\citep{Savrasov_2003}
the second derivatives of the DFT+DMFT functional were calculated
at finite $\mathbf{q}$ vector and neglecting some terms; the work
of Leonov \emph{at al.}\citep{Leonov2014} proposed the forces calculation,
which was not based on a stationary functional and required calculation
of two-particle vertex at all frequencies and implied building an
effective Hubbard model to be solved by the DMFT method.

Recently, a method for analytical calculation of the atomic forces
within DFT+DMFT all-electron linearized augmented plane-wave (LAPW)
formalism was proposed\citep{Haule2016}. Compared to earlier approaches\citep{Savrasov_2003,Leonov2014},
it allowed to derive a general expression for the atomic forces which
is independent on the DMFT solver used. It was shown\citep{Haule2016}
that the use of the total free energy functional at charge self-consistency
greatly simplifies the final expression since several terms cancel
out. The use of all-electron formalism allows to consider only the
standard terms in the Hamiltonian (ion-ion, ion-electron, electron-electron)
which are local. On the other hand, the formalism employing the pseudopotentials, allowing to significantly extend the system size and
capable to calculate the forces within
DFT+DMFT method is still missing. In addition, the use of the
non-orthogonal LAPW basis introduces additional terms into the formalism
and it would be desirable to extend the formalism to a simpler case
of plane-wave basis set. 

Motivated by the above considerations, in
our paper, we show that the formalism developed in Ref.\onlinecite{Haule2016}
can be efficiently extended to a case of both norm-conserving and
ultrasoft pseudopotential DFT, derive all the necessary formulas
and show its benchmark on a real system. The main difficulties outlined
above, will be addressed in details in the subsequent sections.

This paper is organized as follows: in Sections~\ref{sec:Formalism}-\ref{sec:Vanderbilt}
we show how the theory of ultrasoft pseudopotentials can be combined
with the DFT+DMFT formalism and how the atomic forces can be derived
starting from the resulting free-energy functional; in Section~\ref{sec:Results}
we present the benchmark of our formalism on examples of Ce$_{2}$O$_{3}$
and PrO$_2$
and give the conclusions in Section~\ref{sec:Conclusions}. 

\section{Generating functional\label{sec:Formalism}}

The DFT+DMFT total free energy functional was derived in 
Refs.~\onlinecite{Georges2004,Savrasov_2004,Kotliar_2006,Pourovskii2007}
and is reported here for completeness. The starting point is the Baym-Kadanoff
(or Luttinger-Ward) functional (for a review see Ref.~\onlinecite{Kotliar_2006}),
which is a functional of electronic density $\rho(\mathbf{r})$ and
lattice Green function $G_{\nu,\nu^{\prime}}\left(
\mathbf{k},i\omega_{n}\right)\equiv G$:
%with the respective constraint fields $v_{KS}(\mathbf{r})$ and $\Sigma\left(i\omega_{n}\right)$:

\begin{align}
& \Gamma\left[\rho,G\right]=\mathrm{Tr}\ln{G} +U(\mathbf{R})\nonumber \\
 & -\int
 d\mathbf{r}\left(V_{xc}(\mathbf{r})+V_{H}(\mathbf{r})\right)\rho(\mathbf{r})
 -\mathrm{Tr}\left(G^{-1}_0 - G^{-1} \right)G\nonumber \\
\label{eq:GammaAE}\\
 & 
 +E_H[\rho]+E_{xc}[\rho]
 +\sum_{I}\left(\Phi^{\mathrm{DMFT}}[G]-\Phi^{\mathrm{DC}}[G]\right)\nonumber %\\
%&v_{KS}(\mathbf{r})  =V^{ion}(\mathbf{r})+V_{ex}+V_H.\nonumber 
%&\Sigma_1  = \Sigma -V^{DC}.\nonumber
\end{align}
Here, $E_{H}[\rho]$ is the Hartree density functional, $E_{xc}[\rho]$
is the exchange-correlation functional,
 $U(\mathbf{R})$ is the ion-ion Coulomb
interaction energy,
$G_0$ is the DFT Green function:
\[
G_0^{-1} = i\omega_n +\boldsymbol{\mu} - \hat{T} - v_{KS},
\]
where $\boldsymbol{\mu}$ is the system's chemical potential,
$\hat{T}$ is the kinetic energy operator, $v_{KS}$ is the KS
potential:
\[
v_{KS}=V^{ion}+V_{xc}+V_{H},
\]
being $V^{ion}$ the periodic potential of the ions.
$\Phi^{\mathrm{DMFT}}[G]$ is the DMFT interaction
functional, $\Phi^{\mathrm{DC}}[G]$ is the
double-counting functional respectively.
For a detailed discussion of these functionals see Refs.~\onlinecite{Haule_2015,Haule2016}.
The expression~(\ref{eq:GammaAE}) for the $\Gamma$ functional
corresponds to the following expression for the free energy:
\begin{align}
F & =\mathrm{Tr\ln}{G}+E_{H}-\mathbf{\mathrm{Tr}}\left(V_{H}\rho\right)+E_{xc}-\mathrm{\mathbf{\mathrm{Tr}}}\left(V_{xc}\rho\right)\nonumber \\
 & +\sum_{I}
 \left(\Phi^{\mathrm{DMFT}}[G]-\Phi^{\mathrm{DC}}[G]\right)\label{eq:FAE}\\
 & -\mathrm{Tr}G^{loc}\left(\Sigma - V^{DC} \right)+U(\mathbf{R})+\boldsymbol{\mu}\mathcal{N}.\nonumber 
\end{align}
Here, $\mathcal{N}$
is the number of electrons in the unit cell, and the reason why the term $\boldsymbol{\mu}\mathcal{N}$ was added to the free energy expression in the context of force calculation will be explained in the subsequent sections.
$V_{xc}$ and $V_{H}$ are the exchange and Hartree potentials
respectively, while $\Sigma$ is the self-energy and $V^{\mathrm{DC}}$ is
the double-counting potential:
\begin{align}
V_{xc}&=\frac{\delta E_{xc}}{\delta\rho} \\
V_H&=\frac{\delta E_{H}}{\delta\rho} \\
\Sigma &=\frac{\delta\Phi^{\mathrm{DMFT}}[G]}{\delta G} \\ 
V^{\mathrm{DC}}&=\frac{\delta\Phi^{\mathrm{DC}}[G]}{\delta G}.
\end{align}
Finally, $G^{loc}$ is the local Green function which will be defined below.
%

%The symbol $I$ represents the ion index.
%$\Sigma^{B}(\mathbf{r},i\omega_{n})$ is the Bloch self-energy obtained
%by upfolding of $\Sigma_1$ (explained below), while 
The trace operator appearing in the Eqs.~(\ref{eq:GammaAE},\ref{eq:FAE}),
for a general matrix function (or operator) $A$ is defined as:
\begin{equation}
\mathrm{Tr}A=T\sum_{n,l}A_{ll}(i\omega_{n})e^{i\omega_{n}0^{+}},\label{eq:defTr}
\end{equation}
\emph{i.e.} traced over both orbital and imaginary time indices at
temperature $T$. 
%Here, we use the Atomic Hartree units, so that $\hbar=1$,
%$e=1$ and $m_{e}=1$. 

% On the other hand, the variation with respect to $v_{KS}$ and $\Sigma_1$,
%taking into account \eqref{eq:constraints} yields 
The lattice Green function $G$ and the electronic density $\rho$
are obtained respectively as:
\begin{align}
   G(\mathbf{k},i\omega_{n})&=\left(i\omega_{n}+\boldsymbol{\mu}-\hat{T}
   -v_{KS}-\Sigma^{B}(\mathbf{k},i\omega_{n})\right)^{-1}\nonumber \\
\label{eq:min_rho}\\
\rho\left(\mathbf{r}\right) &=\mathrm{Tr}\langle\mathbf{r}\left|\hat{G}\right|\mathbf{r}\rangle\nonumber.
%G_{m,m^{\prime}}^{I} & =\left\langle \beta_{m}^{I}\left|\hat{G}\right|\beta_{m^{\prime}}^{I}\right\rangle ,\nonumber 
\end{align}
Here, $\Sigma^B$ is the
lattice self-energy obtained from $\Sigma$ and $V^{\mathrm{DC}}$ by the
so-called upfolding transformation:
\begin{equation}
    \Sigma^B_{\nu,\nu^\prime}(\mathbf{k},i\omega_n) = \sum_{L,L^\prime}
    P^{\star}_{\nu,L}(\mathbf{k})
    \left(
    \Sigma - V^{\mathrm{DC}}
    \right)_{L,L^\prime}
    P_{L^\prime,\nu^\prime}(\mathbf{k}).
\end{equation}
Within DFT+DMFT, $\Sigma^B$ acquires the $\mathbf{k}$-dependence, unlike the pure DMFT case where
the self-energy is local.
Here, we have implicitly introduced the projectors onto the localized states $\left\{ \beta_{m}^{I}\right\}$:
\begin{equation}
   \label{defproj}
    P_{L,\nu}(\mathbf{k}) = \langle \beta^{I}_{m}|S|\phi_{\mathbf{k},\nu}\rangle=
    \langle \beta_{L}|S|\phi_{\mathbf{k},\nu}\rangle,
\end{equation}
where the index $L$ comprises the atom position $I$ and the orbital
index $m$: $L\equiv\{m,I\}$. The projectors are defined as the overlaps
between localized states $\beta_{L}$ and the KS orbitals $\phi_{\mathbf{k},\nu}$ with a metric $S$, which
takes care of the non-orthogonality of the $\beta$ states, as was pointed
out in the Ref.~\onlinecite{CASTEPDMFT}. It will become evident in the
following section that this matrix is the same $S$ matrix introduced in
the formalism of the ultrasoft pseudopotentials with the same scope.
The opposite operation - downfolding - is required in order to obtain
the local Green function $G^{loc}$ appearing in the Eq.~(\ref{eq:FAE}):
\begin{equation}
    G^{loc}_{L,L^\prime}(\omega) = \sum_{\mathbf{k},\nu,\nu^\prime} 
    P_{L,\nu}(\mathbf{k}) G_{\nu,\nu^\prime} (\mathbf{k},\omega) P^{\star}_{\nu^\prime,L^\prime}(\mathbf{k}).
\end{equation}
We would like to stress that the above formulas were derived for
the all-electron case, as opposed to the pseudopotential case considered
in the present work. As will be shown in the subsequent
Section, in the latter case there appears an additional non-local
density dependent potential appears in the Hamiltonian so that the above
formalism cannot be applied in its present form.
It is the scope of the present paper to adapt the forces formalism
derived in the Ref.~\onlinecite{Haule2016} to the pseudopotential case.

\section{Vanderbilt's formalism\label{sec:Vanderbilt}}

Here, we extend the all-electron DFT+DMFT formalism to the case when the pseudopotentials are used.
Ultra-soft pseudopotentials (USPP) were first proposed in the Refs.~\onlinecite{Vanderbilt1990,Vanderbilt1993}.
The advantage of USSP over the norm-conserving pseudopotentials consists
in lowering the cutoff energy for the plane waves thanks to relaxing
the condition of norm conservation and allowing for non-orthogonality
of the local projectors. The norm-conserving pseudopotentials can
be viewed of as a limiting case of USPP if the norm conservation is
imposed. For what regards force calculation, several difficulties
arise in the case when the pseudopotentials are employed within DFT+DMFT:
i) DFT Hamiltonian contains non-local projection term, which implicitly
depends on the density; ii) the Kohn-Sham (KS) eigenvectors become
non-orthogonal; iii) electronic density contains the augmentation
part in addition to the usual plane-wave one; and iv) USPP method
is formulated by using the total internal energy, while, for the forces
calculation, the total free energy is more preferable. We show below,
how these points can be addressed, and notice, regarding the point
ii) that it was shown in the Ref.\onlinecite{CASTEPDMFT}, how the
non-orthogonality of the local basis within DFT+DMFT can be efficiently
taken into account by using the projection overlap matrix $S$ as
a metric. We start by rewriting the USSP total energy, proceed by
showing that the DFT forces, derived from this expression is identical
to the usual USSP formula and, finally extend the formalism to the
case of DFT+DMFT. 

\subsubsection{Reformulating USSP free energy and forces}
By using the KS eigenvalues:

\begin{align}
\sum_{\mathbf{k},\nu}o_{\mathbf{k},\nu}\varepsilon_{\mathbf{k},\nu} & =\sum_{\mathbf{k},\nu}o_{\mathbf{k},\nu}\left\langle \phi_{\mathbf{k},\nu}\left|-\nabla^{2}+V_{NL}^{(0)}\right|\phi_{\mathbf{k},\nu}\right\rangle \\
 & +\int d\mathbf{r}V_{eff}(\mathbf{r})\rho(\mathbf{r}),\nonumber 
\end{align}

the USSP total energy can be rewritten (at self-consistency) as follows
(in the notations of Ref.\onlinecite{Vanderbilt1993}):

\begin{align}
\label{Etot_vander}
   E_{tot}&=\sum_{\mathbf{k},\nu}o_{\mathbf{k},\nu}\varepsilon_{\mathbf{k},\nu}+E_{H}[\rho]
   -\mathbf{\mathrm{Tr}}\left(V_{H}\rho\right)\\
   \nonumber
   &+E_{xc}[\rho]
   -\mathrm{\mathbf{\mathrm{Tr}}}\left(V_{xc}\rho\right)+U(\mathbf{R}).
\end{align}

Here, $o_{\mathbf{k},\nu}$ is the $\nu$-th KS level occupancy at
momentum $\mathbf{k}$, $V_{NL}^{(0)}$ is the ``unscreened'' non-local
potential, $V_{eff}(\mathbf{r})$ is the effective potential:
\begin{equation}
V_{eff}(\mathbf{r})=V^{ion}(\mathbf{r})+V_{H}(\mathbf{r})+V_{xc}(\mathbf{r}).
\end{equation}
Finally, $U(\mathbf{R})$ is the interatomic Coulomb interaction energy
and $\rho$ represents the full electronic charge density (plane wave
plus augmentation).

Variating $E_{tot}$ with respect to an atomic position $\mathbf{R}_{\mu}$,
we obtain:
\begin{align}
\mathbf{\mathbf{F}}_{\mu} & =-\frac{\partial E_{tot}}{\partial\mathbf{R}_{\mu}}\label{eq:Etotvar}\\
 & =-\sum_{\mathbf{k},\nu}o_{\mathbf{k},\nu}\frac{\delta\varepsilon_{\mathbf{k},\nu}}{\delta\mathbf{R}_{\mu}}+\mathbf{\mathrm{Tr}}\left(\frac{\delta\left(V_{H}+V_{xc}\right)}{\delta\mathbf{R}_{\mu}}\rho\right)-\frac{\partial U}{\partial\mathbf{R}_{\mu}}.\nonumber 
\end{align}
$\frac{\delta\varepsilon_{\mathbf{k},\nu}}{\delta\mathbf{R}_{\mu}}$
can be easily obtained from the Schrodinger equation by using the
Hellmann-Feynman theorem: 

\begin{align*}
H\left|\phi_{\mathbf{k},\nu}\right\rangle  & =\varepsilon_{\mathbf{k},\nu}S\left|\phi_{\mathbf{k},\nu}\right\rangle \\
\frac{\delta\varepsilon_{\mathbf{k},\nu}}{\delta\mathbf{R}_{\mu}} & =\left\langle \phi_{\mathbf{k},\nu}\left|\frac{\delta H}{\delta\mathbf{R}_{\mu}}\right|\phi_{\mathbf{k},\nu}\right\rangle -\varepsilon_{\mathbf{k},\nu}\left\langle \phi_{\mathbf{k},\nu}\left|\frac{\delta S}{\delta\mathbf{R}_{\mu}}\right|\phi_{\mathbf{k},\nu}\right\rangle .
\end{align*}
Here, $H$ is the effective (non physical) Hamiltonian defined with
the ``screened'' non-local part as:
\begin{align}
H & =-\nabla^{2}+V_{NL}+V_{eff}(\mathbf{r}),\\
\label{defS}
S & =1+\sum_{n,m,I}q_{nm}\left|\beta_{n}^{I}\right\rangle \left\langle \beta_{m}^{I}\right|,
\end{align}
with $V_{NL}$ - self-consistent non-local projection operator:
\begin{equation}
V_{NL}=\sum_{n,m,I}D_{nm}^{I}\left|\beta_{n}^{I}\right\rangle \left\langle \beta_{m}^{I}\right|,
\end{equation}
as opposed to the ''bare`` non-local projectors:
\begin{equation}
   V^{(0)}_{NL}=\sum_{n,m,I}D_{nm}^{(0)}\left|\beta_{n}^{I}\right\rangle \left\langle\beta_{m}^{I}\right|.
\end{equation}
$D_{nm}^{I}$ and $D_{nm}^{(0)}$ are connected through the charge augmentation:
\begin{equation}
   \label{defDnm}
D_{nm}^{I}=D_{nm}^{(0)}+\int d\mathbf{r}V_{eff}(\mathbf{r})Q_{nm}^{I}(\mathbf{r}).
\end{equation}
Here, the quantities $D_{nm}^{(0)}$ and $Q_{nm}^I(\mathbf{r})$ are the properties of the
pseudopotential, as explained in the Ref.~\onlinecite{Vanderbilt1993}, and $D_{nm}^{(0)}$
does not change when variating the atomic positions.
The local functions $\beta_n^I$ are also part the pseudopotential definition, 
although they are centered at the ions and do move rigidly with the atoms. The matrix $S$ is the cause of non-orthogonality of the KS eigenvectors. 

In Eq.\eqref{eq:Etotvar}, we neglected the variation of $o_{\mathbf{k},\nu}$
because within the DFT USPP formalism the forces calculations are carried out at zero temperature, and
the occupancies are assumed to be step-function-like. Below, within DFT+DMFT formalism, the
variation of DMFT occupancies will be shown to cancel out if the forces are derived from the total
free energy.
%in the spirit of Ref.\onlinecite{Haule2016}.

Recording that:
\[
\frac{\delta H}{\delta\mathbf{R}_{\mu}}=\frac{\delta V_{eff}(\mathbf{r})}{\delta\mathbf{R}_{\mu}}+\frac{\delta V_{NL}}{\delta\mathbf{R}_{\mu}}
\]
and after some simplifications, we get:\begin{widetext}
\begin{equation}
   \label{dHdR}
\sum_{\mathbf{k},\nu}o_{\mathbf{k},\nu}\left\langle \phi_{\mathbf{k},\nu}\left|\frac{\delta H}{\delta\mathbf{R}_{\mu}}\right|\phi_{\mathbf{k},\nu}\right\rangle =\mathbf{\mathrm{Tr}}\left(\frac{\delta V_{eff}}{\delta\mathbf{R}_{\mu}}\rho\right)+\sum_{n,m,I}\int d\mathbf{r}V_{eff}(\mathbf{r})\frac{\partial Q_{nm}^{I}(\mathbf{r})}{\partial\mathbf{R}_{\mu}}\rho_{nm}^{I}+\sum_{n,m,I}D_{nm}^{I}\frac{\partial\rho_{nm}^{I}}{\partial\mathbf{R}_{\mu}}.
\end{equation}
\end{widetext}

   Here we have used the following properties: i) the definition of $D_{nm}^{I}$
   (Eq.~(\ref{defDnm})), ii) the fact that $\delta D_{nm}^{(0)}/\delta \mathbf{R}_\mu=0$; iii) the
   definitions of the full density $\rho(\mathbf{r})$, the quantity $\rho^I_{nm}$ and its derivative
   $\frac{\partial \rho^I_{nm}}{\partial \mathbf{R}_\mu}$ from the
   Ref.~\onlinecite{Vanderbilt1993}:
\begin{align}
   \rho^I_{nm} &= \sum_{\mathbf{k},\nu}o_{\mathbf{k},\nu}\left\langle \phi_{\mathbf{k},\nu}\left|\beta_{n}^{I}\right\rangle
   \left\langle \beta_{m}^{I}\right|\phi_{\mathbf{k},\nu}\right\rangle \\
   \rho(\mathbf{r}) &= \sum_{\mathbf{k},\nu}o_{\mathbf{k},\nu}\left| \phi_{\mathbf{k},\nu}(\mathbf{r})\right|^2
   +\sum_{n,m,I}Q^I_{nm}(\mathbf{r}) \rho^I_{nm}\\
   \frac{\partial \rho^I_{nm}}{\partial \mathbf{R}_\mu} &=
   \sum_{\mathbf{k},\nu}o_{\mathbf{k},\nu} \left[
   \left\langle \phi_{\mathbf{k},\nu}\left|\frac{\partial \beta_{n}^{I}}{\partial \mathbf{R}_\mu}\right\rangle
   \left\langle \beta_{m}^{I}\right|\phi_{\mathbf{k},\nu}\right\rangle\right.\\
   &+
   \left.\left\langle \phi_{\mathbf{k},\nu}\left|\beta_{n}^{I}\right\rangle
   \left\langle \frac{\partial \beta_{m}^{I}}{\partial \mathbf{R}_\mu}\right|\phi_{\mathbf{k},\nu}\right\rangle
   \right].
   \nonumber
\end{align}
With these definitions, it is easy to derive the Eq.~(\ref{dHdR}).

On the other hand, the metrics part (containing the
derivative of $S$) becomes:
\[
\sum_{\mathbf{k},\nu}o_{\mathbf{k},\nu}\varepsilon_{\mathbf{k},\nu}\left\langle
\phi_{\mathbf{k},\nu}\left|\frac{\delta
   S}{\delta\mathbf{R}_{\mu}}\right|\phi_{\mathbf{k},\nu}\right\rangle=\sum_{n,m,I}q_{nm}\frac{\partial\omega_{nm}^{I}}{\partial\mathbf{R}_{\mu}}.
\]

Here, once again, we have used the definitions of $\omega_{nm}^{I}$ and $\frac{\partial \omega^I_{nm}}{\partial \mathbf{R}_\mu}$ from the
Ref.~\onlinecite{Vanderbilt1993} with $\Lambda_{\mathbf{k},\nu;\mathbf{k^{\prime}},\nu^{\prime}}=\varepsilon_{\mathbf{k},\nu}\delta_{\mathbf{k},\nu;\mathbf{k}^{\prime},\nu^{\prime}}$, which corresponds to the equilibrium condition, as explained therein:
\begin{align}
   \omega^I_{nm} &= \sum_{\mathbf{k},\nu}o_{\mathbf{k},\nu}\varepsilon_{\mathbf{k},\nu}
   \left\langle \phi_{\mathbf{k},\nu}\left|\beta_{n}^{I}\right\rangle
   \left\langle \beta_{m}^{I}\right|\phi_{\mathbf{k},\nu}\right\rangle \\
   \frac{\partial \omega^I_{nm}}{\partial \mathbf{R}_\mu} &=
   \sum_{\mathbf{k},\nu}o_{\mathbf{k},\nu}\varepsilon_{\mathbf{k},\nu} \left[
   \left\langle \phi_{\mathbf{k},\nu}\left|\frac{\partial \beta_{n}^{I}}{\partial \mathbf{R}_\mu}\right\rangle
   \left\langle \beta_{m}^{I}\right|\phi_{\mathbf{k},\nu}\right\rangle\right.\\
   &+
   \left.\left\langle \phi_{\mathbf{k},\nu}\left|\beta_{n}^{I}\right\rangle
   \left\langle \frac{\partial \beta_{m}^{I}}{\partial \mathbf{R}_\mu}\right|\phi_{\mathbf{k},\nu}\right\rangle
   \right].
   \nonumber
\end{align}

Putting all the terms together, we, indeed, obtain the standard USPP
force formula (see Ref.~\onlinecite{Vanderbilt1993}):

\begin{widetext}
\begin{equation}
\mathbf{F}_{\mu}=-\frac{\partial U}{\partial\mathbf{R}_{\mu}}-\mathbf{\mathrm{Tr}}\left(\frac{\delta V^{ion}}{\delta\mathbf{R}_{\mu}}\rho\right)-\sum_{n,m,I}\int d\mathbf{r}V_{eff}(\mathbf{r})\frac{\partial Q_{nm}^{I}(\mathbf{r})}{\partial\mathbf{R}_{\mu}}\rho_{nm}^{I}-\sum_{n,m,I}D_{nm}^{I}\frac{\partial\rho_{nm}^{I}}{\partial\mathbf{R}_{\mu}}+\sum_{n,m,I}q_{nm}\frac{\partial\omega_{nm}^{I}}{\partial\mathbf{R}_{\mu}}.\label{eq:FDFT}
\end{equation}
\end{widetext}
This expression is identical to the Eq.(43) of the Ref.~\onlinecite{Vanderbilt1993}.

\subsubsection{Formulating USSP DFT+DMFT free energy}
Now, we turn to the Eq.~\ref{Etot_vander}. We can easily generalize
it to the DFT+DMFT case and write directly the generating functional
$\Gamma$ and the free energy $F$:
\begin{align}
\Gamma\left[G\right] & =\mathrm{Tr\ln}\hat{G}-\mathrm{Tr}\left(\left\{ \hat{G}_{0}^{-1}-\hat{G}^{-1}\right\} \hat{G}\right)+E_{H}+E_{xc}\label{eq:GammaPP}\\
 & +\sum_{I}\Phi^{\mathrm{DMFT}}[G]-\sum_{I}\Phi^{\mathrm{DC}}[G]+U(\mathbf{R})\nonumber 
\end{align}

\begin{align}
F & =\mathrm{Tr\ln}\hat{G}+E_{H}-\mathbf{\mathrm{Tr}}\left(V_{H}\rho\right)+E_{xc}-\mathrm{\mathbf{\mathrm{Tr}}}\left(V_{xc}\rho\right)\nonumber \\
 & +\sum_{I}\Phi^{\mathrm{DMFT}}[G]-\sum_{I}\Phi^{\mathrm{DC}}[G]-\mathrm{Tr}\left(\left(\Sigma-V^{\mathrm{DC}}\right)G\right)\label{eq:FPP}\\
 & +U(\mathbf{R})+\boldsymbol{\mu}\mathcal{N}.\nonumber 
\end{align}
In passing from $\Gamma\left[G\right]$ to $F$ the following expression
for $\hat{G}$ was obtained:
\begin{align*}
\hat{G}(\mathbf{k},i\omega_n) & =\left[i\omega_{n}+\boldsymbol{\mu}-\varepsilon_{\mathbf{k},\nu}-\Sigma^{B}(\mathbf{k},i\omega_n)\right]^{-1},
\end{align*}
where by definition $\varepsilon_{\mathbf{k},\nu}=E_{kin}+V^{ion}+V_{NL}+V_{H}+V_{xc}$
in KS basis, and we cast $G_{0}$ (Green function in the absence of
$\Sigma$ and $V_{\mathrm{DC}}$ - DFT Green function):
\[
\hat{G}_{0}^{-1}=i\omega_{n}+\boldsymbol{\mathbf{\mathbf{\mathbf{\mathbf{\mu}}}}}-T-V^{ion}-V_{NL}-V_{H}-V_{xc}=i\omega_{n}+\boldsymbol{\mu}-\varepsilon_{\mathbf{k},\nu}.
\]
Here, $E_{kin}$ is the electron's kinetic energy, $\boldsymbol{\mu}$ is the system's chemical potential, $\mathcal{N}$
is the number of electrons in the unit cell. The reason why $+\boldsymbol{\mu}\mathcal{N}$
term is added is because the free energy is defined as: $F=E-T\mathbf{S}-\boldsymbol{\mu}\mathcal{N}$
with $\mathbf{S}$ as the system's entropy, and we do not want the
$-\boldsymbol{\mu}\mathcal{N}$ term to contribute to the forces.
Comparing Eqs.\eqref{eq:GammaAE}-\eqref{eq:FAE}
and Eqs.\eqref{eq:GammaPP}-\eqref{eq:FPP}, we can see that the non-local
projection term can be absorbed into the definitions of $G$ and $G_{0}$
so that the final expressions for $\Gamma$ and $F$ are identical
to those of the all-electron DFT+DMFT. In addition, we note that there holds a Dyson equation in Bloch space:

\begin{align}
    \hat{G}^{-1}(\mathbf{k},i\omega_n) = \hat{G}_0^{-1}(\mathbf{k},i\omega_n)-
    \Sigma^B(\mathbf{k},i\omega_n).
\end{align}

%For the following forces derivation, we express all the matrices
%in the KS basis since the beginning, as we expect a great simplification
%of the formulas.

Now, we can check the limiting case of DFT forces by deriving them
directly from $F$:
\begin{align*}
F^{DFT} & =\mathrm{Tr\ln}\hat{G}+E_{H}-\mathbf{\mathrm{Tr}}\left(V_{H}\rho\right)\\
 & +E_{xc}-\mathrm{\mathbf{\mathrm{Tr}}}\left(V_{xc}\rho\right)+U(\mathbf{R})+\boldsymbol{\mu}\mathcal{N}.
\end{align*}
In the DFT case, obviously, $H^{KS}$ expressed in the KS basis is
a diagonal matrix with the corresponding eigenvalues $\varepsilon_{\mathbf{k},\nu}$
on the diagonal. Variating with respect to an ionic coordinate $\mathbf{R}_{\mu}$,
we obtain:
\begin{align*}
\mathbf{F}_{\mu} & =-\mathbf{\mathrm{Tr}}\left(G\frac{\delta\varepsilon_{\mathbf{k},\nu}-\delta\mu}{\delta\mathbf{R}_{\mu}}\right)+\mathbf{\mathrm{Tr}}\left(\frac{\delta\left(V_{H}+V_{xc}\right)}{\delta\mathbf{R}_{\mu}}\rho\right)\\
 & -\frac{\partial U}{\partial\mathbf{R}_{\mu}}-\mathcal{N}\frac{\delta\boldsymbol{\mu}}{\delta\mathbf{R}_{\mu}}.
\end{align*}
Once again, in the DFT case, $\varepsilon_{\mathbf{k},\nu}$ does
not depend on $\omega$ and, hence, the sum on $\omega$ part of the
trace can be done, giving:
\begin{align*}
\mathbf{\mathrm{Tr}}\left(G\frac{\delta}{\delta\mathbf{R}_{\mu}}\left(\varepsilon_{\mathbf{k},\nu}-\boldsymbol{\mu}\right)\right) & =\sum_{\mathbf{k},\nu}o_{\mathbf{k},\nu}\frac{\delta}{\delta\mathbf{R}_{\mu}}\left(\varepsilon_{\mathbf{k},\nu}-\boldsymbol{\mu}\right)\\
 & =\sum_{\mathbf{k},\nu}o_{\mathbf{k},\nu}\frac{\delta\varepsilon_{\mathbf{k},\nu}}{\delta\mathbf{R}_{\mu}}-\mathcal{N}\frac{\delta\boldsymbol{\mu}}{\delta\mathbf{R}_{\mu}}.
\end{align*}
Putting everything together, we obtain:
\[
\mathbf{F}_{\mu}^{DFT}=-\sum_{\mathbf{k},\nu}o_{\mathbf{k},\nu}\frac{\delta\varepsilon_{\mathbf{k},\nu}}{\delta\mathbf{R}_{\mu}}+\mathbf{\mathrm{Tr}}\left(\frac{\delta\left(V_{H}+V_{xc}\right)}{\delta \mathbf{R}_{\mu}}\rho\right)-\frac{\partial U}{\partial\mathbf{R}_{\mu}},
\]
which is identical to Eq.\eqref{eq:Etotvar}. Here, the occupancies
are defined according to the definition \eqref{eq:defTr} (except
for the omitted summation on $\nu$) as: $o_{\mathbf{k},\nu}=\mathrm{Tr}G_{\nu,\nu}(\mathbf{k},i\omega_{n})$. 

Let us see how the number of particles is calculated in the Vanderbilt's
pseudopotential formalism:

\begin{widetext}

\begin{align*}
\mathcal{N} & =\int\rho(\mathbf{r})d\mathbf{r}=\sum_{\mathbf{k},\nu}o_{\mathbf{k},\nu}\int d\mathbf{r}\left\{ \left|\phi_{\mathbf{k},\nu}(\mathbf{r})\right|^{2}+\sum_{n,m}Q_{n,m}(\mathbf{r})\left\langle \phi_{\mathbf{k},\nu}|\beta_{n}^{I}\right\rangle \left\langle \beta_{m}^{I}|\phi_{\mathbf{k},\nu}\right\rangle \right\} \\
 & =\sum_{\mathbf{k},\nu}o_{\mathbf{k},\nu}\left\{ \left\langle \phi_{\mathbf{k},\nu}|\phi_{\mathbf{k},\nu}\right\rangle +\sum_{n,m}q_{n,m}\left\langle \phi_{\mathbf{k},\nu}|\beta_{n}^{I}\right\rangle \left\langle \beta_{m}^{I}|\phi_{\mathbf{k},\nu}\right\rangle \right\} \\
 & =\sum_{\mathbf{k},\nu}o_{\mathbf{k},\nu}\left\{ \left\langle \phi_{\mathbf{k},\nu}|\phi_{\mathbf{k},\nu}\right\rangle +\left\langle \phi_{\mathbf{k},\nu}\left|S-1\right|\phi_{\mathbf{k},\nu}\right\rangle \right\} =\sum_{\mathbf{k},\nu}o_{\mathbf{k},\nu}\left\langle \phi_{\mathbf{k},\nu}\left|S\right|\phi_{\mathbf{k},\nu}\right\rangle =\sum_{\mathbf{k},\nu}o_{\mathbf{k},\nu}.
\end{align*}
\end{widetext}

Here we used the fact that $q_{n,m}=\int Q_{n,m}(\mathbf{r})d\mathbf{r}$
and the definition of $S$ from Ref.~\onlinecite{Vanderbilt1993}. 

\subsubsection{USSP DFT+DMFT forces}

Variating with respect to $\mathbf{R}_{\mu}$, and using the above definitions, we obtain:

\begin{align}
\mathbf{F}_{\mu}^{\mathrm{DMFT}} & =-\mathrm{Tr}\sum_{\mathbf{k},\nu,\nu^{\prime}}\frac{\delta\widetilde{\varepsilon}_{\nu\nu^{\prime}}(\mathbf{k},i\omega_{n})}{\delta\mathbf{R}_{\mu}}G_{\nu^{\prime}\nu}(\mathbf{k},i\omega_{n})\nonumber \\
 & +\mathbf{\mathrm{Tr}}\left(\rho\frac{\delta}{\delta\mathbf{R}_{\mu}}\left(V_{H}+V_{xc}\right)\right)-\frac{\partial U}{\partial\mathbf{R}_{\mu}}\label{eq:FDMFT}\\
 & +\mathbf{\mathrm{Tr}}\left(G^{loc}\frac{\delta}{\delta\mathbf{R}_{\mu}}\left(\Sigma-V^{\mathrm{DC}}\right)\right),\nonumber 
\end{align}
where $\widetilde{\varepsilon}_{\nu\nu^{\prime}}(\mathbf{k},i\omega_{n})\equiv\varepsilon_{\mathbf{k},\nu}\delta_{\nu\nu^{\prime}}+\Sigma_{\nu\nu^{\prime}}^{B}(\mathbf{k},i\omega_{n})$ and the Green function, density and self-energy are expressed in the KS basis.

Therefore, 
\begin{alignat*}{1}
 & \frac{\delta\widetilde{\varepsilon}_{\nu\nu^{\prime}}(k,i\omega_{n})}{\delta\mathbf{R}_{\mu}}=\delta_{\nu,\nu^{\prime}}\frac{\delta\varepsilon_{k,\nu}}{\delta\mathbf{R}_{\mu}}+\frac{\delta \Sigma^{B}}{\delta\mathbf{R}_{\mu}}\\
 & =\delta_{\nu,\nu^{\prime}}\left\langle \phi_{\mathbf{k},\nu}\left|\frac{\delta H}{\delta\mathbf{R}_{\mu}}\right|\phi_{\mathbf{k},\nu}\right\rangle -\delta_{\nu,\nu^{\prime}}\varepsilon_{\mathbf{k},\nu}\left\langle \phi_{\mathbf{k},\nu}\left|\frac{\delta S}{\delta\mathbf{R}_{\mu}}\right|\phi_{\mathbf{k},\nu}\right\rangle \\
 & +\frac{\delta P_{\nu,L}^{\star}(\mathbf{k})}{\delta\mathbf{R}_{\mu}}\left(\Sigma-V^{\mathrm{DC}}\right)_{L,L^{\prime}}P_{L^{\prime},\nu^{\prime}}(\mathbf{k})\\
 & +P_{\nu,L}^{\star}(\mathbf{k})\left(\Sigma-V^{\mathrm{DC}}\right)_{L,L^{\prime}}\frac{\delta P_{L^{\prime},\nu^{\prime}}(\mathbf{k})}{\delta\mathbf{R}_{\mu}}\\
 & +P_{\nu,L}^{\star}(\mathbf{k})\frac{\delta}{\delta\mathbf{R}_{\mu}}\left(\Sigma-V^{\mathrm{DC}}\right)_{L,L^{\prime}}P_{L^{\prime},\nu^{\prime}}(\mathbf{k}).
\end{alignat*}
The last term in this expression, when substituted into $\mathbf{F}_{\mu}^{\mathrm{DMFT}}$
cancels out the last term in Eq.\eqref{eq:FDMFT}, and we note that
the first line, involving $\frac{\delta H}{\delta R_{\mu}}$ and $\frac{\delta S}{\delta R_{\mu}}$
is independent on frequency, so that the trace on omega can be evaluated,
giving the DMFT occupancy:
\[
o_{\mathbf{k},\nu}^{\mathrm{DMFT}}=\mathrm{Tr}G_{\nu,\nu}(\mathbf{k},i\omega_{n}).
\]
Moreover, the expression:
\[
\sum_{\mathbf{k},\nu}o_{\mathbf{k},\nu}^{\mathrm{DMFT}}\left\{ \left\langle \phi_{\mathbf{k},\nu}\left|\frac{\delta H}{\delta\mathbf{R}_{\mu}}\right|\phi_{\mathbf{k},\nu}\right\rangle -\varepsilon_{\mathbf{k},\nu}\left\langle \phi_{\mathbf{k},\nu}\left|\frac{\delta S}{\delta\mathbf{R}_{\mu}}\right|\phi_{\mathbf{k},\nu}\right\rangle \right\} 
\]
has the same functional form as in the Vanderbilt's theory of USPP and can be brought into the form of Eq.\eqref{eq:FDFT}
where $o_{\mathbf{k},\nu}$ is substituted with $o_{\mathbf{k},\nu}^{\mathrm{DMFT}}$.
In doing that, we have to remember that in Eq.\eqref{eq:FDFT} the
terms $\mathbf{\mathrm{Tr}}\left(\rho\frac{\delta}{\delta\mathbf{R}_{\mu}}\left(V_{H}+V_{xc}\right)\right)-\frac{\partial U}{\partial\mathbf{R}_{\mu}}$
are already taken into account and, in particular, the former is partially
cancelled out, leaving the $\mathbf{\mathrm{-Tr}}\left(\frac{\delta V^{ion}}{\delta\mathbf{R}_{\mu}}\rho\right)$
term.

The final formula for the DFT+DMFT forces can be expressed as follows in analogy with the Ref.~\onlinecite{Haule2016}:
\begin{equation}
\mathbf{F}_{\mu}^{\mathrm{DMFT}}=\widetilde{\mathbf{F}}_{\mu}^{DFT}+\mathbf{F}_{\mu}^{dyn},\label{eq:FDMFT-final}
\end{equation}
where $\widetilde{\mathbf{F}}_{\mu}^{DFT}$ is the force, calculated
according to Eq.~\eqref{eq:FDFT} with the occupancy $o_{\mathbf{k}\nu}^{\mathrm{DMFT}}$
instead of $o_{\mathbf{k}\nu}^{DFT}$ in the total density $\rho(\mathbf{r})$
(shown below) and in the following expressions (that is why ``tilde''):
\begin{align*}
\widetilde{\rho}_{nm}^{I} & =\sum_{\mathbf{k},\nu}o_{\mathbf{k},\nu}^{\mathrm{DMFT}}\left\langle \phi_{\mathbf{k},\nu}|\beta_{n}^{I}\right\rangle \left\langle \beta_{m}^{I}|\phi_{\mathbf{k},\nu}\right\rangle \\
\widetilde{\omega}_{nm}^{I} & =\sum_{\mathbf{k},\nu}o_{\mathbf{k},\nu}^{\mathrm{DMFT}}\left\langle \phi_{\mathbf{k},\nu}|\beta_{n}^{I}\right\rangle \left\langle \beta_{m}^{I}|\phi_{\mathbf{k},\nu}\right\rangle \varepsilon_{\mathbf{k},\nu}.
\end{align*}
Now, the full charge self-consistency DFT+DMFT implies:

\begin{align*}
\rho(\mathbf{r}) & =\sum_{\mathbf{k},\nu}o_{\mathbf{k},\nu}^{\mathrm{DMFT}}\left\{ \left|\phi_{\mathbf{k},\nu}(\mathbf{r})\right|^{2}+\sum_{n.m}Q_{n,m}^{I}(\mathbf{r})\left\langle \phi_{\mathbf{k},\nu}|\beta_{n}^{I}\right\rangle \left\langle \beta_{m}^{I}|\phi_{\mathbf{k},\nu}\right\rangle \right\} \\
 & =\sum_{\mathbf{k},\nu}o_{\mathbf{k},\nu}^{\mathrm{DMFT}}\left|\phi_{\mathbf{k},\nu}(\mathbf{r})\right|^{2}+\sum_{n,m}Q_{n,m}^{I}(\mathbf{r})\widetilde{\rho}_{nm}^{I}.
\end{align*}
On the other hand, $V_{eff}$, depending on the full electronic density
$\rho(\mathbf{r})$ and entering into $\widetilde{\mathbf{F}}_{\mu}^{DFT}$
explicitly and through $D_{n,m}^{I}$ has to be taken at the ``self-consistency'', as was pointed out in the Ref.~\onlinecite{Haule2016}.

$\mathbf{F}_{\mu}^{dyn}$ can be expressed as:

\begin{widetext}

\begin{alignat}{1}
\mathbf{F}_{\mu}^{dyn} & =-\mathrm{Tr}\sum_{\mathbf{\substack{\mathbf{k},\nu,\nu^{\prime}\\
\mathit{L,L^{\prime}}
}
}}\left\{ \frac{\delta P_{\nu,L}^{\star}(\mathbf{k})}{\delta\mathbf{R}_{\mu}}\left(\Sigma-V^{\mathrm{DC}}\right)_{L,L^{\prime}}P_{L^{\prime},\nu^{\prime}}(\mathbf{k})+P_{\nu,L}^{\star}(\mathbf{k})\left(\Sigma-V^{\mathrm{DC}}\right)_{L,L^{\prime}}\frac{\delta P_{L^{\prime},\nu^{\prime}}(\mathbf{k})}{\delta\mathbf{R}_{\mu}}\right\} G_{\nu^{\prime}\nu}(\mathbf{k},i\omega_{n})\nonumber \\
 & =-\mathrm{Tr}\sum_{L,L^{\prime}}\left(\Sigma(i\omega_{n})-V^{\mathrm{DC}}\right)_{L,L^{\prime}}\sum_{\mathbf{k},\nu,\nu^{\prime}}\left\{ P_{L^{\prime},\nu^{\prime}}(\mathbf{k})G_{\nu^{\prime}\nu}(\mathbf{k},i\omega_{n})\frac{\delta P_{\nu,L}^{\star}(\mathbf{k})}{\delta\mathbf{R}_{\mu}}+\frac{\delta P_{L^{\prime},\nu^{\prime}}(\mathbf{k})}{\delta\mathbf{R}_{\mu}}G_{\nu^{\prime}\nu}(\mathbf{k},i\omega_{n})P_{\nu,L}^{\star}(\mathbf{k})\right\} \nonumber \\
 & =-\mathrm{Tr}\sum_{L,L^{\prime}}\left(\Sigma(i\omega_{n})-V^{\mathrm{DC}}\right)_{L,L^{\prime}}\Xi_{L^{\prime},L}(i\omega_{n}).\label{eq:Fdyn}
\end{alignat}
where we have defined the function $\Xi$:
\begin{equation}
\Xi_{L^{\prime},L}(i\omega_{n})=\sum_{\mathbf{k},\nu,\nu^{\prime}}\left\{ P_{L^{\prime},\nu^{\prime}}(\mathbf{k})G_{\nu^{\prime}\nu}(\mathbf{k},i\omega_{n})\frac{\delta P_{\nu,L}^{\star}(\mathbf{k})}{\delta\mathbf{R}_{\mu}}+\frac{\delta P_{L^{\prime},\nu^{\prime}}(\mathbf{k})}{\delta\mathbf{R}_{\mu}}G_{\nu^{\prime}\nu}(\mathbf{k},i\omega_{n})P_{\nu,L}^{\star}(\mathbf{k})\right\} .\label{eq:Xi}
\end{equation}

\end{widetext}

The use of the time reversal symmetry in the numerical evaluation of the Matsubara sums is
exemplified in the Appendix~\ref{appb}.

\subsubsection{Derivation of the projectors derivatives}

In this subsection, we summarize the formulas necessary to calculate the derivatives of the
projectors to the localized states $P_{L,\nu}(\mathbf{k})$. From the definition Eq.(\ref{defproj}) we have:
\begin{align}
   &\frac{\delta P_{L,\nu}(\mathbf{k})}{ \delta \mathbf{R}_\mu} =
   \frac{\delta \langle \beta_L \left| S \right| \phi_{\mathbf{k},\nu}\rangle}{ \delta
	  \mathbf{R}_\mu} =\\
   &\langle \frac{\delta \beta_L}{ \delta \mathbf{R}_\mu} \left| S \right|
   \phi_{\mathbf{k},\nu}\rangle +
   \langle \beta_L \left| \frac{\delta S}{ \delta \mathbf{R}_\mu} \right| \phi_{\mathbf{k},\nu}\rangle,
\end{align}
where:
%we have taken into account the Hellman-Feynman theorem according to which:
\begin{equation}
   \frac{\delta }{\delta \mathbf{R}_\mu} \left| \phi_{\mathbf{k},\nu}\rangle \right.=0,
\end{equation}
since the KS orbitals do not depend explicitly on atomic coordinates~\citep{martin_2004,Cococcioni2014,Cococcioni2020}.
%(The KS orbitals do not depend explicitly on atomic coordinates, or,
%equivalently, the eigenstates do not change at the first order of the
%perturbation theory~\citep{martin_2004,Cococcioni2014,Cococcioni2020}).
The derivative of $S$ can be readily calculated, starting from the definition Eq.(\ref{defS}):
\begin{equation}
   \frac{\delta S}{ \delta \mathbf{R}_\mu} = \sum_{n,m,I} q_{nm}
   \left(
	  \left| \frac{\delta \beta_n^I}{\delta \mathbf{R}_\mu} \rangle \right.
	  \left. \langle  \beta_m^I\right| +
	  \left| \beta_n^I  \rangle \right.
	  \left. \langle  \frac{\delta\beta_m^I}{\delta \mathbf{R}_\mu} \right|
   \right).
\end{equation}
At this point, we would like to recall that the objects $q_{nm}$ and $\left| \beta_n^I  \rangle \right.$
are determined at the pseudopotential generation stage and remain unchanged
during DFT+DMFT density optimization. The only dependence on $\mathbf{R}_\mu$ in $\left| \beta_n^I  \rangle \right.$
comes from the fact that these localized orbitals move rigidly with their corresponding ions, so
that the derivatives $\left| \frac{\delta \beta_n^I}{\delta \mathbf{R}_\mu} \rangle \right.$ can be
calculated by going into momentum representation, exactly as it is done in the
Refs.~\onlinecite{Vanderbilt1990,Vanderbilt1993} or in the Ref.~\onlinecite{Haule2016}.

\section{Benchmarks and results\label{sec:Results}}

\subsection{Forces in cerium sesquioxide}

\begin{figure*}
\includegraphics[width=1\textwidth]{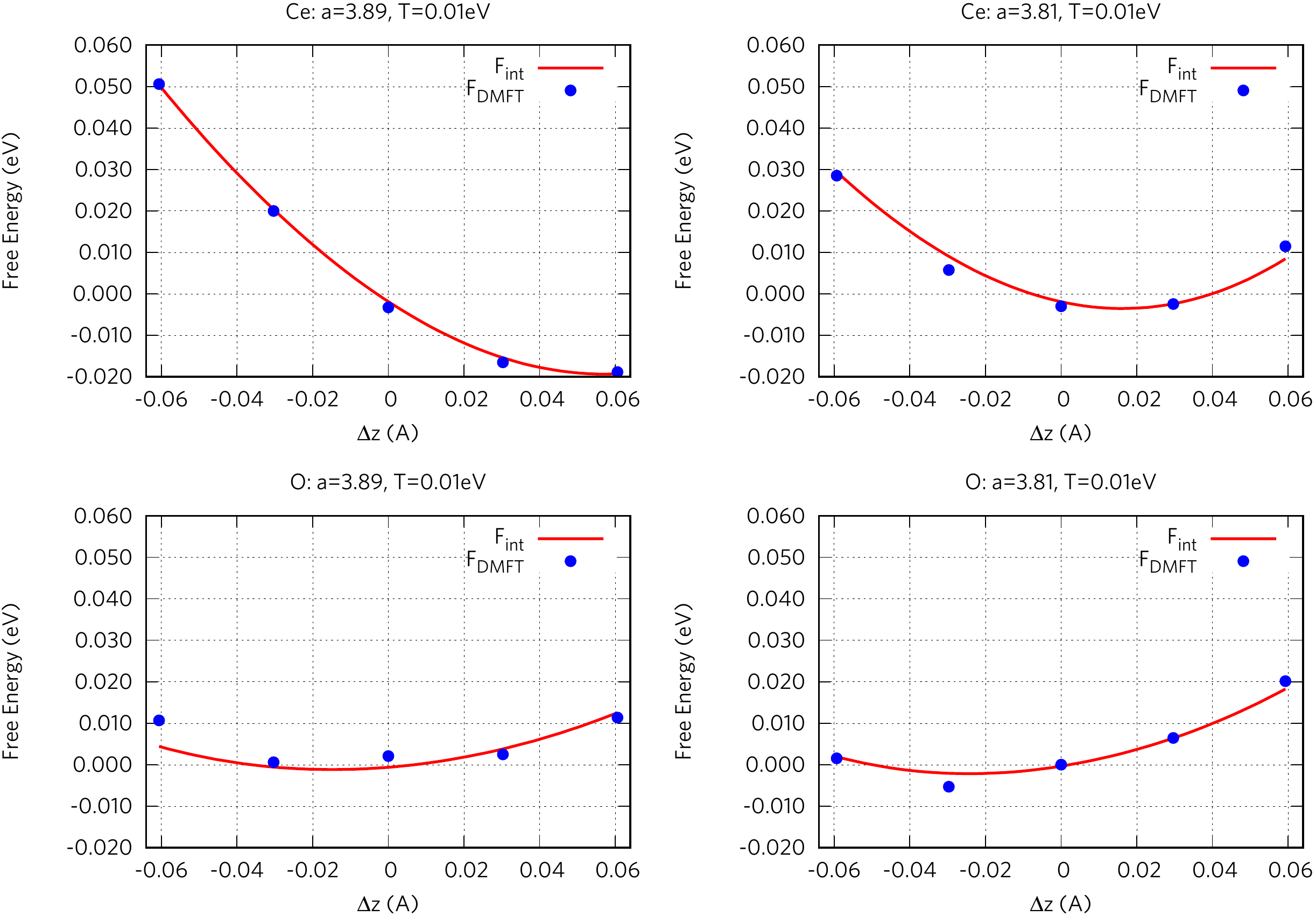}

\caption{\label{fig:Energies}Energy profiles (blue points) in Ce$_{2}$O$_{3}$
when displacing Ce (top row) and O (bottom row) along $z$ direction.
Left column corresponds to $a=3.89$ \AA, while right column corresponds
to $a=3.81$\AA. The red curves correspond to the free energy profiles
derived from integrating the analytical DFT+DMFT forces.}
\end{figure*}

\begin{figure*}
\includegraphics[width=1\textwidth]{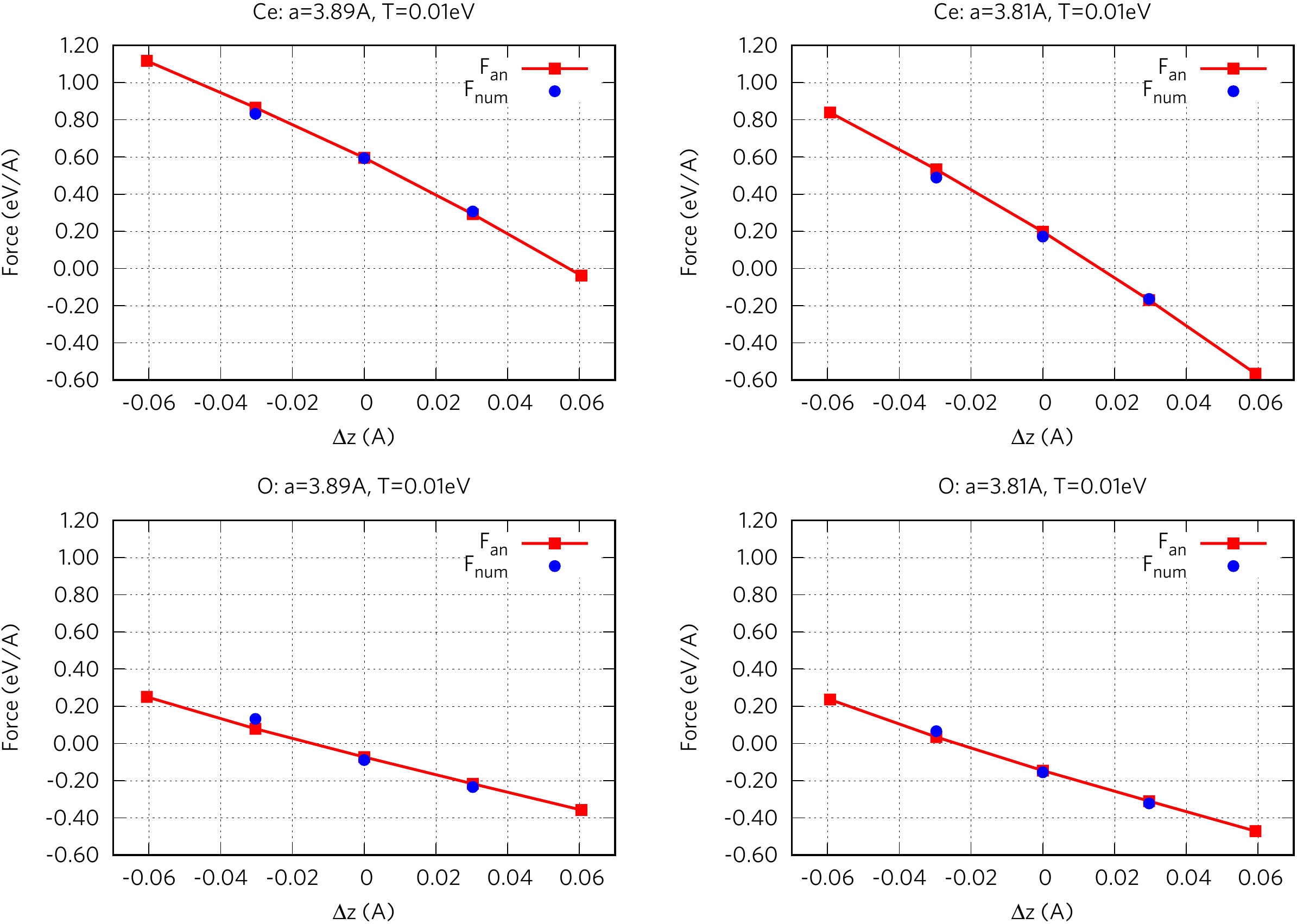}

\caption{\label{fig:Forces}Comparison of the DFT+DMFT forces. Red points:
analytical forces; blue points: numerical derivative extracted from
the numerical free energy profiles reported in Fig.\ref{fig:Energies}.
Forces acting on Ce (top row) and O (bottom row) along $z$ direction.
Left column corresponds to $a=3.89$ \AA, while right column corresponds
to $a=3.81$ \AA.}
\end{figure*}

The results presented in this section are obtained by implementing
the formulas presented above within the DFT+DMFT method implemented
previously\citep{CASTEPDMFT,CTQMCCASTEP} in the widely used plane-wave
DFT code CASTEP\citep{RMP-Payne,CASTEP}. In order to benchmark our
formalism, we apply it to Cerium sesquioxide Ce$_{2}$O$_{3}$, which
has been studied for a long time\citep{Andersson_2007,Fabris_2005,Singh_2006,Loschen_2007}.
It is known to be an anti-ferromagnetic insulator with N\'eel temperature
of $T_{N}=9$~K and a gap of $2.4$~eV. DFT+DMFT calculations in the
literature normally address the high-temperature paramagnetic phase,
so to benchmark our forces calculations we also set the temperature
to $T=0.01$~eV. Ce$_{2}$O$_{3}$ crystallizes in a hexagonal unit
cell with space group $P\bar{3}m1$. The experimental parameters for
the unit cell are: $a=3.89$ \AA\ and $c/a=1.557$, with the Wyckoff
positions\citep{Wyckoff67}: Ce $2d$ $\left(\frac{1}{3},\frac{2}{3},\xi\right)$,
O $2d$ $(\frac{1}{3},\frac{2}{3},\eta)$, O $1a$ $(0,0,0)$, with
$\xi=0.24543$ and $\eta=0.6471$. On the other hand DFT predicts
$a=3.81$ \AA$\;$ at the experimental ratio $c/a=1.557$ and experimental
$\xi$ and $\eta$. We have performed calculations for both lattice
constants $a=3.81$ \AA$\;$(minimum energy for DFT+DMFT method) and
$a=3.89$ \AA$\;$(the experimental value), while maintaining the
ratio $c/a=1.557$. We have used the norm-conserving Ce and O pseudopotential
(NCP17 set), LDA exchange-correlation potential and a $31\times31\times17$
Monkhorst-Pack $k$-point mesh. We have also checked that similar
results are obtained with the ultrasoft pseudopotentials too. The
plane-wave basis cut-off was automatically determined to be $1012$ eV.
The values of Hubbard $U$ and Hund $J$ parameters were chosen to be $U=6$ eV and $J=0.7$ eV
respectively.
The results for Ce$_{2}$O$_{3}$ density of states at the experimental
geometry are shown our previous work\citep{CASTEPDMFT} and exhibit
excellent agreement with the reference calculations of Ref.~\onlinecite{Pourovskii2007}.
As in our previous paper, the DMFT calculations were performed with
the Hubbard I solver
(the extensions to the other types of solvers e.g. Hubbard III~\cite{Mizia2009,Mizia2011} can be done)
with a fixed occupancy of $n=1$ per Ce atom
(in the sense explained in Ref.~\onlinecite{Pourovskii2007}) within
the FLL double-counting scheme.

For the benchmark to be fair, we compare the analytical forces calculated
within our formalism against the numerical ones obtained from finite
increment derivative of the free energy. 
On the other hand, we also compare the numerical free energy profiles
against the curves obtained from the spline integration of the analytical
forces. For what regards the evaluation of the numerical forces, we
first note that most internal atomic coordinates are fixed by symmetry.
We vary the remaining coordinates, which are the $z$-coordinates
of Ce $2d$ and O $2d$ atoms (the ones established from experiment).
Obviously, the forces of the atoms related by symmetry are in turn
related. During finite increment of relevant atomic coordinates, we
tested several $\Delta z$ values, in order to be sure that the free 
energy varies linearly over the lengthscale of $\Delta z$. The results
of these tests are shown in Fig.8 of Ref.\onlinecite{CASTEPDMFT},
and in this work we fix $\Delta z=1\%$ in units of the $c$-dimension
of the unit cell. The numerical forces were determined as follows:
\begin{equation}
   \mathbf{F}_{z_{i}}=-\frac{\partial F_{tot}}{\partial z_{i}}.\label{eq:Fnum}
\end{equation}
In addition, we emphasize that the total free energy as a function of $\Delta z$
is a smooth differentiable function, thanks to the fact that both
DFT (CASTEP) and DMFT subsystems in our calculations are well-behaved,
giving small responses to small perturbations. In order to be consistent
with the formalism developed in the previous Section, in the present
work, the DFT+DMFT was self-consistently converged until the energy
became stationary up to $10^{-6}$eV.

The comparison between the analytical forces, calculated within the
formalism presented in the previous section, and the numerical forces,
derived from the total free energy according to Eq.~\eqref{eq:Fnum}, is
shown in Figs.~\ref{fig:Energies},\ref{fig:Forces}. The energy profiles
are presented in Fig.~\ref{fig:Energies}, while the forces comparison
is illustrated in Fig.~\ref{fig:Forces}. The overall agreement appears
to be very good, taking into account the inevitable numerical bias
of the DFT+DMFT total free energy. The forces calculated within our formalism
are correct for both Ce - correlated ion, and O - \textquotedbl uncorrelated
ion\textquotedbl , on which the dynamical force is identically zero.
We note that the local minimum (where the force is zero) with respect
to the Ce displacement along $z$-axis is approximately $+0.06$ \AA\; 
with respect to the experimental position for the $a=3.89$ \AA\; unit
cell, while it is about $+0.017$ \AA\; for the $a=3.81$ \AA\; unit cell.
In the case of O displacement, the order of magnitude of forces is
smaller, while the minima positions are roughly $-0.02$ \AA\; for both
unit cells considered. Compared to the DFT forces (Table II of Ref.~\onlinecite{CASTEPDMFT}),
the Ce DFT+DMFT forces presented here are larger, while the O forces
are smaller. Compared to the one-shot DFT+DMFT forces (same table
of Ref.~\onlinecite{CASTEPDMFT}), the full charge self-consistency
modifies significantly the resulting force: for Ce - increasing, while
for O decreasing. We conclude, therefore, that the one-shot DFT+DMFT
somehow overshoots the forces with respect to the full self-consistent
DFT+DMFT. 
It was shown in the Ref.~\onlinecite{Pourovskii2007}, that
the full self-consistent DFT+DMFT gives somewhat better agreement
with the experiment for the Ce$_2$O$_3$ equilibrium volume as compared to the one-shot DFT+DMFT,
thanks to the spectral weight redistribution.
In addition, the difference between the DFT and the DMFT forces is larger
on the \textquotedbl correlated\textquotedbl\;ions, although the
\textquotedbl uncorrelated\textquotedbl\;ones are also modified
due to the fact that the density is distributed differently in DFT+DMFT
with respect to DFT. On the other hand, we have checked that the total
vector sum of all the forces acting on all the atoms in
the unit cell is zero within both DFT and DFT+DMFT as it should be in equilibrium. 

\subsection{Forces in praseodymium dioxide}

\begin{figure}
\includegraphics[width=1\columnwidth]{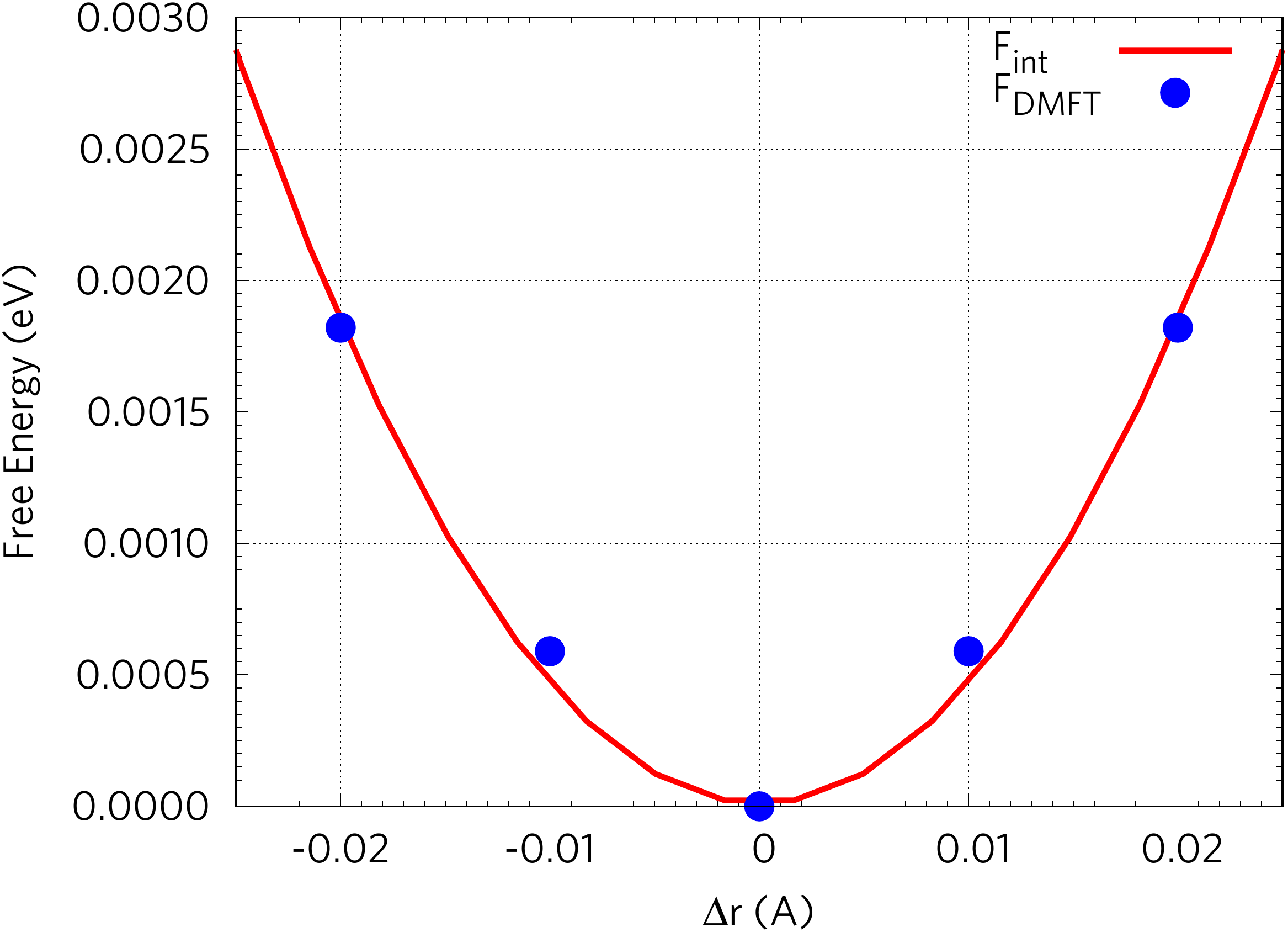}

\caption{\label{fig:Energy_Pr}
Energy profiles (blue points) in PrO$_{2}$
when displacing Pr atom along the $(111)$ direction.
$\mathrm{F_{int}}$ stands for the spline-integrated forces along the displacement
path, while $\mathrm{F_{DMFT}}$ stands for system's free energy calculated at
given atomic positions using full charge self-consistency.
}
\end{figure}

In order to enforce the validity of our approach, we have benchmarked the DMFT
forces in yet another system: praseodymium dioxide PrO$_2$.
We consider PrO$_2$ in the rhombohedral unit cell (symmetry
group $Fm\bar{3}m$) with $a=4.0482$ \AA\; and the following Wyckoff
positions of the atoms: Pr at $(0,0,0)$ and two oxygen atoms at
$(\frac{1}{4},\frac{1}{4},\frac{1}{4})$ and
$(\frac{3}{4},\frac{3}{4},\frac{3}{4})$
respectively~\citep{Chiba2011,Andreeva1986}. Here, we have used
ultra-soft pseudopotentials for both Pr and O (C17 set), LDA
exchange-correlation potential and a $25\times25\times25$
Monkhorst-Pack $k$-point mesh. The plane-wave basis cut-off was
automatically determined to be $653$ eV.
The values of Hubbard $U$ and Hund $J$ parameters were chosen to be $U=6$ eV and $J=0.7$ eV
respectively.
At the above Wyckoff
positions, the net DFT+DMFT forces are zero due to symmetry and the
finite forces appear if the corresponding atoms are pushed away from
their positions. Since both Pr and O atoms are placed on the cubic
cell diagonal, in doing the finite displacements it is important
to conserve the $3$-fold axis along the diagonal. That is why in the
present subsection, we
have performed the finite displacements of the Pr atom along the
$(111)$ direction.
The free energy increment between two atomic positions $\mathbf{R}^1$
and $\mathbf{R}^2$ has then been estimated by using the
following formula:
\begin{equation}
   F_{tot}(\mathbf{R}^2) - F_{tot}(\mathbf{R}^1) =
   -\int_{\mathbf{R}^1}^{\mathbf{R}^2} \sum_{\mu}
   \mathbf{F}_{\mu}(\mathbf{R}) \mathrm{d} \mathbf{R}_{\mu},
\end{equation}
where $\mathbf{F}_{\mu}(\mathbf{R})$ is the $\mu$'s component of the force
at the atomic position vector $\mathbf{R}$, while $\mathbf{R}_{\mu}$
is the $\mu$'s Cartesian coordinate of the displaced Pr atom. 
The excellent agreement between $\delta F_{tot}$ derived from the analytical
forces and the free energy profiles calculated in the vicinity of the
high-symmetry Wyckoff position of Pr atom is shown in
Fig.~\ref{fig:Energy_Pr}.
The forces appear to be
symmetric with respect to the displacements of the atoms along the
diagonal in the positive and negative directions off the exact
Wyckoff positions and so does the free energy profile.
We would like to point out that in the case
of PrO$_2$ the order of magnitude of forces and energy increments
associated with the atomic displacements are an order of
magnitude smaller as compared to the Ce$_2$O$_3$ case, which required
additional accuracy in deriving smooth free energy profile.

\section{\label{sec:Conclusions}Conclusions}

In conclusion, we have presented a formalism for analytic calculation
of the atomic forces within the full charge self-consistent pseudopotential
DFT+DMFT approach. Our approach extends that of the Ref.~\onlinecite{Haule2016}
by taking into account the non-local projections terms in the KS Hamiltonian,
which depend implicitly on charge distribution and arise from the
pseudization procedure. It inherits the useful properties of the DFT+Embedded
DMFT functional\citep{Haule2016} and, in particular $\frac{\delta P}{\delta G}=0$,
and, therefore, the terms most difficult to calculate cancel out from
the final result. Plane-wave basis, employed within our implementation,
greatly simplifies the formalism by avoiding calculation of the augmentation
charges. Our formalism is implemented within the DMFT framework inside
CASTEP \emph{ab-initio} code, which in the past already allowed for
precise total free energy calculations within DFT+DMFT\citep{CASTEPDMFT}.
Our approach is general and suitable for both norm-conserving and
ultrasoft pseudopotentials.
The pseudopotential approach has an advantage of speeding up the calculations with respect to the
all-electron methods by considering the core electrons as frozen, while the ultra-soft
pseudopotentials further speed up the calculations with
respect to the norm-conserving pseudopotentials by relaxing the norm-conserving
condition~\cite{Vanderbilt1990,Vanderbilt1993}.

In addition, our approach does not use any specific
DMFT solver property and, hence, 
would work 
equally well with all solvers.
We have presented the benchmark of our approach on the example of
Ce$_{2}$O$_{3}$, which showed excellent agreement between the forces
analytically calculated within our approach and the forces obtained
from numerical differentiation of the total free energy at very low temperature.
In addition, we have compared the total free energy profiles against the
integrated forces profiles which also showed excellent agreement.
We analyzed the differences of atomic forces within DFT, one-shot
DFT+DMFT and full charge self-consistent DFT+DMFT on the 
examples of
Ce$_{2}$O$_{3}$ and PrO$_2$,
the applicability to the correlated metal close to the Mott transition being subject of our future
studies.
Our approach allows for quick and reliable force
calculations within fully self-consistent pseudopotential DFT+DMFT
and paves the way to the structural optimization, phonon and molecular
dynamics calculations within DFT+DMFT. 
\begin{acknowledgments}
This work was performed using resources provided by the Cambridge
Service for Data Driven Discovery (CSD3) operated by the University
of Cambridge Research Computing Service (\url{www.csd3.cam.ac.uk}),
provided by Dell EMC and Intel using Tier-2 funding from the Engineering
and Physical Sciences Research Council (capital grant EP/P020259/1),
and DiRAC funding from the Science and Technology Facilities Council
(\url{www.dirac.ac.uk}), Project cs085. In addition, this work used
the computational support from the Cirrus UK National Tier-2 HPC Service
at EPCC (\url{http://www.cirrus.ac.uk}) funded by the University
of Edinburgh and EPSRC (EP/P020267/1), Project ec130. 
\end{acknowledgments}

\appendix

\section{Calculation of the free energy}
We start from Eq.~(\ref{eq:FPP}) for the total free energy. 
We also report for completeness the formula for the total internal energy, used e.g. in our past
work~\cite{CASTEPDMFT} (taking into account the ion-ion interaction energy):
\begin{align}
E & =\sum \varepsilon_{\mathbf{k},\nu} N_{\nu,\nu}(\mathbf{k})+E_{H}-\mathbf{\mathrm{Tr}}\left(V_{H}\rho\right)+E_{xc}-\mathrm{\mathbf{\mathrm{Tr}}}\left(V_{xc}\rho\right)\nonumber \\
 & +\frac{1}{2} \mathrm{Tr} \Sigma G-\sum_{I}\Phi^{\mathrm{DC}}[G]+U(\mathbf{R}),\label{eq:EPP}
\end{align}
where $N_{\nu,\nu}(\mathbf{k})=T\sum_{n} G(\mathbf{k},i\omega_n)$ is the DMFT occupancy matrix and
$\varepsilon_{\mathbf{k},\nu}$ are the DFT eigenvalues calculated at the DMFT density.
With respect to the total internal energy calculation, the
changes are the following:
\begin{itemize}
   \item The term $\sum \varepsilon_{\mathbf{k},\nu} N_{\nu,\nu}(\mathbf{k})$ is substituted with
	  the following expression:
	  \begin{equation}
		 \mathrm{Tr}\ln \hat{G} +\boldsymbol{\mu}\mathcal{N} - \mathrm{Tr} \left(
			\Sigma - V^{DC}
		 \right) G;
	  \end{equation}
   \item The term $\frac{1}{2} \mathrm{Tr} \Sigma G$ is substituted with:
	  \begin{equation}
		 \Phi^{\mathrm{DMFT}} = F_{imp} - \mathrm{Tr}\ln G_{imp} + \mathrm{Tr}
			\Sigma_{imp} G_{imp}.
	  \end{equation}
\end{itemize}
The calculation of the $\mathrm{Tr}\ln G$ with a general Green function $G$ is performed following
the procedure outlined in the Ref.~\onlinecite{Haule_2015_Free_Energy}, namely: the summation is split into
two parts - numerical sum up to a cutoff Matsubara frequency $i\omega_c$ with the most divergent
part subtracted, and an expression equal to the known analytical sum of the most divergent part. In
this case, the most divergent part is:
\begin{equation}
   \left. -T \sum_n \ln(-i\omega_n + \varepsilon ) e^{i\eta \omega_n}\right|_{\eta\to 0} = 
   -T \ln\left(1 + e^{-\frac{\varepsilon}{T}}\right).
\end{equation}
Therefore, the summation $\mathrm{Tr}\ln \hat{G}$ is evaluated as follows:
\begin{widetext}
\begin{align}
   \mathrm{Tr}\ln \hat{G} &= -T\sum_{\substack{ |\omega_n|<\omega_c \\ \mathbf{k}}} 
   \mathfrak{Tr}
   \left\{
	 \ln\left( -i\omega_n + \varepsilon_{\mathbf{k},\nu}
		- \boldsymbol{\mu} + \Sigma^{B}_{\nu,\nu^\prime}(\mathbf{k},i\omega_n) - V_{\nu,\nu^\prime}^{DC} \right) -
	 \ln\left(-i\omega_n + \varepsilon_{\nu,\nu^\prime}(\mathbf{k},\infty) \right)
  \right\} \\
  &- T \sum_{\mathbf{k}} \ln
  \left(1 + \exp\left({-\frac{\varepsilon_{\nu,\nu^\prime}(\mathbf{k},\infty)}{T}}\right)\right),
\end{align}
\end{widetext}
where $\varepsilon_{\nu,\nu^\prime}(\mathbf{k},\infty)=\varepsilon_{\mathbf{k},\nu}
- \boldsymbol{\mu} + \Sigma^{B}_{\nu,\nu^\prime}(\mathbf{k},\infty)-V_{\nu,\nu^\prime}^{DC}$.

On the other hand, the
summation $\mathrm{Tr}\ln G_{imp}$ is evaluated as:
\begin{widetext}
\begin{align}
   \mathrm{Tr}\ln G_{imp} &= -T\sum_{|\omega_n|<\omega_c} 
   \mathfrak{Tr}
   \left\{
	  \ln\left( -i\omega_n + \varepsilon^{imp}_{m,m^{\prime}}
		 + \Sigma^{imp}_{m,m^{\prime}}(i\omega_n)\right) -
	  \ln\left(-i\omega_n + \varepsilon^{imp}_{m,m^{\prime}}(\infty) \right)
   \right\} \\
   &- T \ln\left(1 + \exp\left(-\frac{\varepsilon^{imp}_{m,m^{\prime}}(\infty)}{T}\right)\right),
\end{align}
\end{widetext}
where this time $\varepsilon^{imp}_{m,m^{\prime}}(\infty)=
\varepsilon^{imp}_{m,m^{\prime}}+\Sigma^{imp}_{m,m^{\prime}}(\infty)$. Here, the notation
$\mathfrak{Tr}$ stands for the trace over $\nu,\nu^\prime$ or $m,m^\prime$ indices respectively
(without the summation over Matsubara frequencies).

\section{\label{appb} The use of the time reversal symmetry in the Matsubara sums}

When doing the sums like $-\mathrm{Tr}\Sigma(i\omega_{n})G(i\omega_{n})$,
one usually makes use of the symmetry properties of $\Sigma$ and
$G$ upon changing $i\omega_{n}\to-i\omega_{n}$:
\begin{align*}
G_{m,m^{\prime}}(-i\omega_{n}) & =G_{m^{\prime},m}^{\star}(i\omega_{n})\\
\Sigma_{m,m^{\prime}}(-i\omega_{n}) & =\Sigma_{m^{\prime},m}^{\star}(i\omega_{n}),
\end{align*}
so that 
\begin{align*}
 & -T\sum_{n,m,m^{\prime}}\Sigma_{m,m^{\prime}}(-i\omega_{n})G_{m^{\prime},m}(-i\omega_{n})=\\
 & -T\sum_{n,m,m^{\prime}}G_{m,m^{\prime}}^{\star}(i\omega_{n})\Sigma_{m^{\prime},m}^{\star}(i\omega_{n})
\end{align*}
hence:
\begin{flalign*}
 & -T\sum_{n,m,m^{\prime}}G_{m,m^{\prime}}(i\omega_{n})\Sigma_{m^{\prime},m}(i\omega_{n})=\\
 & -2T\mathrm{Re}\sum_{m,m^{\prime},\omega_{n}>0}G_{m,m^{\prime}}(i\omega_{n})\Sigma_{m^{\prime},m}(i\omega_{n}).
\end{flalign*}
Considering the definition of $\Xi(i\omega_{n})$ given by the Eq.~\eqref{eq:Xi},
we see that indeed:
\[
\Xi_{m,m^{\prime}}(-i\omega_{n})=\Xi_{m^{\prime},m}^{\star}(i\omega_{n}).
\]
Therefore, one can still use the Green function's symmetry properties
and restrict the summation in Eq.~\eqref{eq:Fdyn} to the positive
Matsubara frequencies, while the final formula for the DFT+DMFT forces
is given by the Eq.~\eqref{eq:FDMFT-final}.

\bibliographystyle{apsrev4-1}
\bibliography{biblio}
\end{document}